\documentclass[hf]{ceurart}
\usepackage{tikz}
\usepackage{listings}
\usepackage{graphicx} 
\usepackage{comment} 
\usepackage{pifont}
\usepackage{booktabs}
\usepackage{xcolor}
\usepackage{hyperref}
\usepackage{cleveref}
\usepackage{todonotes}

\newcommand{\fullcircle}{%
  \tikz[baseline=-0.6ex]\fill (0,0) circle (0.6ex);%
}

\newcommand{\emptycircle}{%
  \tikz[baseline=-0.6ex]\draw (0,0) circle (0.6ex);%
}

\newcommand{\halfcircle}{%
  \tikz[baseline=-0.6ex]{
    \draw (0,0) circle (0.6ex);
    \begin{scope}
      \clip (0,-0.6ex) rectangle (0.6ex,0.6ex);
      \fill (0,0) circle (0.6ex);
    \end{scope}
  }%
}

\newcommand{\nb}[3]{
  \fcolorbox{black}{#3}{\color{white}{\bfseries\sffamily\scriptsize#1}}
  {\sf\small\textcolor{#3}{\textit{#2}}}
 }

\newcommand\ivan[2][orange]{\nb{Ivan}{#2}{#1}}

\begin{document}

\copyrightyear{2026}
\copyrightclause{Copyright for this paper by its authors.
  Use permitted under Creative Commons License Attribution 4.0
  International (CC BY 4.0).}

\conference{BP-Meet-IoT 2026: International Workshop on Business Processes Meet the Internet of Things, September 15, 2026, Twente, The Netherlands.}


\title{Emergent Behavior and Uncertainty in IoT-Enhanced Business Processes: Challenges and Future Directions}

\author[1]{Marco Pegoraro}[%
orcid=0000-0002-8997-7517,
email=pegoraro@pads.rwth-aachen.de
]
\address[1]{RWTH Aachen University, Faculty of Computer Science, Aachen, Germany}

\author[2]{Sara Pettinari}[%
orcid=0000-0002-5548-9806,
email=sara.pettinari@gssi.it
]
\author[2]{Ivan Compagnucci}[%
orcid=0000-0002-1991-0579,
email=ivan.compagnucci@gssi.it
]
\address[2]{Gran Sasso Science Institute,
  L'Aquila, Italy}

\author[3]{Marco Franceschetti}[%
orcid=0000-0001-7030-282X,
email=marco.franceschetti@unisg.ch
]
  
\address[3]{University of St.Gallen, Institute of Computer Science, St.Gallen, Switzerland}

\author[1]{Ronny Seiger}[%
orcid=0000-0003-1675-2592,
email=ronny.seiger@rwth-aachen.de,
]

\begin{abstract}
IoT-enhanced business processes are characterized by high complexity due to heterogeneous actors, varying levels of autonomy among participating systems, continuously evolving execution contexts spanning the digital and physical worlds, and continuous event streams. In such settings, process behavior partially emerges only at runtime through complex interactions involving humans, IoT devices, physical objects, software systems, agents, and services. This complexity introduces partial observability, uncertainty, and runtime dynamics that are difficult to anticipate and that challenge traditional business process management (BPM) assumptions and systems. We discuss these challenges from three perspectives, addressing 1)~uncertainty representation, 2)~operationalization of IoT-enhanced processes, and 3)~runtime management of emergent behavior. Based on a motivating scenario and an analysis of the state of the art, we identify open research gaps and outline short-, medium-, and long-term recommendations to shape a research agenda on emergent behavior in IoT-enhanced business processes.
\end{abstract}


\maketitle

\section{Introduction}

IoT-enhanced business processes involve continuous interaction among process instances, physical devices (e.g., sensors, actuators), software services, and intelligent agents--creating an interplay between the physical and digital worlds~\cite{janiesch2020internet}. This interaction makes process behavior more dynamic and less predictable than in traditional BPM settings because execution may depend on uncertain data, changing environmental conditions, real-time events, and autonomous decisions made by IoT components. Nowadays, more and more IoT systems can operate autonomously within their environments to provide useful services. When coming together in unanticipated ways, these self-contained \emph{constituent} systems form \emph{Systems-of-Systems} (SoS) exhibiting \emph{emergent behavior} that is more than the sum of the individual parts~\cite{kopetz2016emergence}. This behavior can be classified as \emph{expected} or \emph{unexpected}, and, more importantly, as having \emph{positive} or \emph{negative} effects, where the latter should be avoided~\cite{perez2014uncertainties}. The main challenge is therefore to move beyond static, centrally controlled process models and to define approaches to modeling, execution, and runtime management that can represent uncertainty, support dynamic behavior, accommodate different interactions among systems, processes, and IoT devices, and respond to runtime changes~\cite{janiesch2020internet}.

The relevance of this problem is evident across domains in which business execution is closely tied to the physical environment through IoT/cyber-physical systems (e.g., smart manufacturing, logistics, healthcare, and smart cities). In these settings, process performance and correctness depend on the ability to monitor real-world events, adapt execution decisions, and dynamically coordinate heterogeneous IoT components in case of unanticipated situations and uncertainties. Addressing this problem requires mechanisms for representing uncertainties,
monitoring and controlling process execution via self-adaptive models and digital twins at runtime, as well as reusable, open-source extensions that make BPM systems more flexible, context-aware, and suitable for IoT-integrated scenarios~\cite{fornari2025digital}.

In this work, we elaborate on the emerging challenges in the development and implementation of evolving IoT-enhanced processes that must become increasingly dynamic and capable of coping with uncertainties in the context of SoS. Our discussion of managing emergent behavior and uncertainties thereby addresses the modeling, operationalization, and runtime phases of IoT-enhanced business processes. With these discussions, we pave the way for future research on IoT-enhanced business processes in real-world domains (e.g., distributed smart manufacturing, smart logistics, smart cities) that require, on the one hand, structured process-based control and analysis of complex systems, and, on the other hand, increasing autonomy of constituent systems to cope with environmental uncertainties during process executions.

The paper is structured as follows: Section~\ref{sec:scenario} presents a motivating scenario. Section~\ref{sec:subchallenges} decomposes the main challenge of managing emergence in IoT-enhanced business processes into three sub-challenges. Section~\ref{sec:sota} discusses the state of the art. Section~\ref{sec:gapsopportunities} elaborates on identified gaps and opportunities. Section~\ref{sec:recommendations} presents our recommendations for future research activities. Section~\ref{sec:conclusions} concludes the paper.

\section{Motivating Scenario: Sources of Uncertainty} \label{sec:scenario}

\begin{figure}
    \centering    \includegraphics[width=0.9\linewidth]{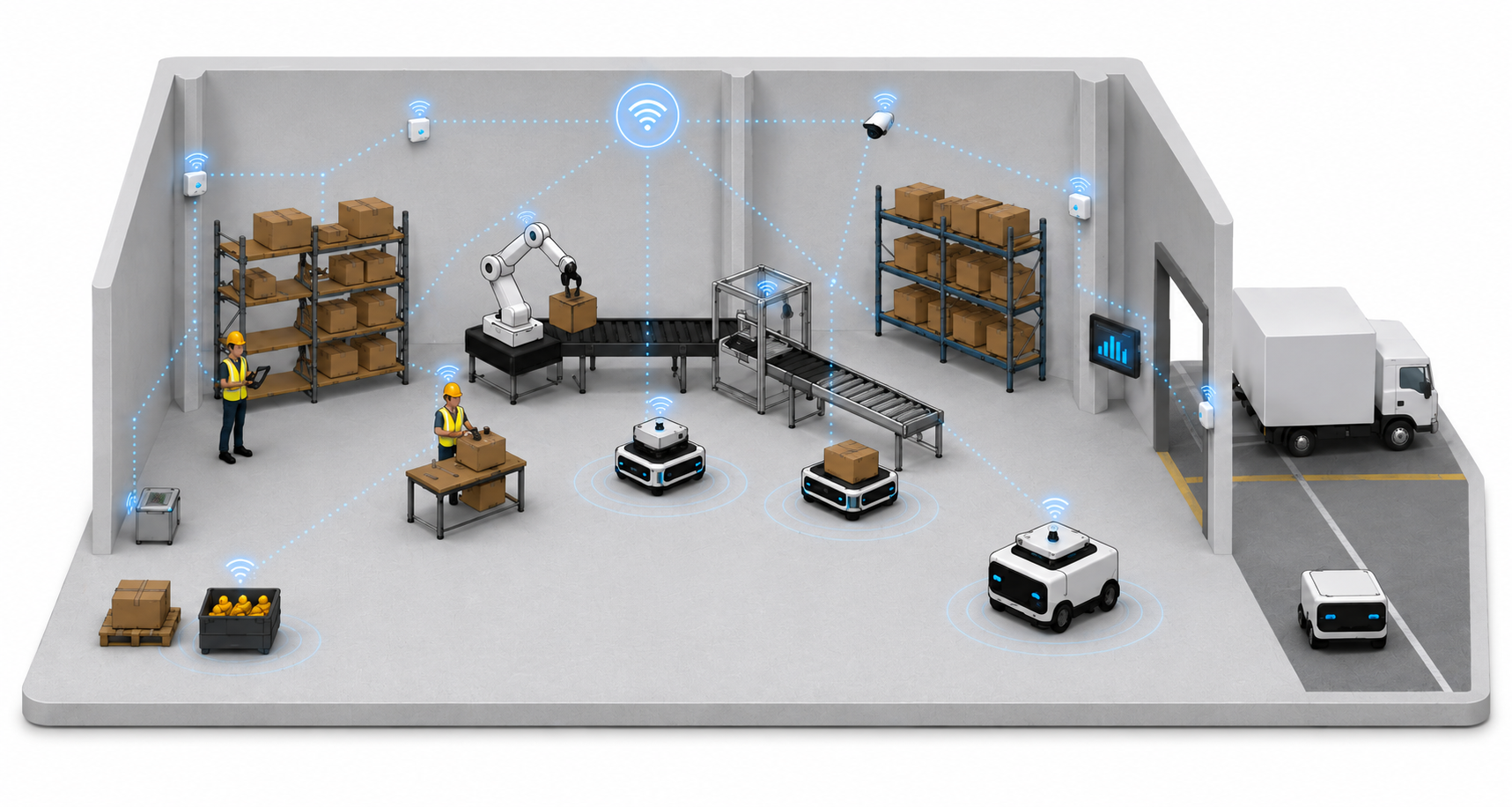}
    \caption{Motivating scenario: emergent smart warehouse operations from the intertwined execution of multiple IoT-enhanced processes by heterogeneous actors. Figure generated using DALL-E by OpenAI.}
    \label{fig:scenario}
\end{figure}

To ground the discussion, we consider a representative IoT-enhanced business process scenario in a \textit{smart warehouse} that involves the collaboration of heterogeneous actors, including human workers, robotic systems, digital information systems, software services, agents, and interconnected IoT devices. Fig.~\ref{fig:scenario} illustrates this scenario. Rather than being orchestrated by a single, well-defined process and process management system, warehouse operations emerge from several intertwined IoT-enhanced processes and ad-hoc interactions that unfold at runtime. Several BPM systems and other information systems that manage only their local execution contexts might be involved. The processes may leverage different IoT interaction modes--active, hybrid, and passive--and must independently adapt to changes in the process context, which may be only partially known to all process participants. Moreover, process actors continuously influence physical and environmental conditions through their close interaction with the real world.

In addition to the internal operations, the scenario extends beyond the warehouse organizational boundaries. For example, delivery-related activities may require interacting with external actors. In such cases, we must account for inherent uncertainty and only partial a priori knowledge about external actors' processes. These external actors (e.g., delivery robots) might be orchestrated by their own processes and operated by different companies or organizations. Thus, they might be constrained by security and privacy restrictions on the process knowledge and on the context they share. Moreover, external entities (e.g.,~humans on the warehouse floor) that are not explicitly modeled within individual IoT-enhanced processes can still influence the emergent process's execution, thereby introducing additional complexity. This can happen because these entities modify the context (e.g.,~the physical environment) in which the process unfolds. For example, an external delivery robot may position itself within the operating area of a robot loading items onto a conveyor belt, thereby obstructing its movement and disrupting its normal operation.

Here, uncertainty emerges from several characteristics of IoT-enhanced business processes: 
\begin{itemize}
    \item Many participating systems are closed and self-managed, exposing only limited interfaces and information while pursuing their own internal goals and missions with their own strategies in a local context. Therefore, their behavior cannot be (fully) anticipated, observed, or controlled by a single process owner. A central entity possessing a complete global view and knowledge of all processes and context is not available.
    \item The participating (constituent) systems are frequently highly heterogeneous and involve devices from different vendors with proprietary logic and decision mechanisms that cannot be exposed, e.g., for privacy reasons. This results in only partial views of the implementation of the participating processes and executing entities.
    \item The presence of actors such as humans that act outside of process control introduces additional uncertainty and variability, e.g., via spontaneous interventions, unforeseen interactions, and unpredictable decisions. These characteristics lead to emergent behavior that cannot be fully modeled a priori, nor can it be addressed through post-mortem analyses (e.g., process mining). Instead, it must be handled dynamically at process runtime -- which is especially important in safety-critical settings where harm or damage must be avoided. 
\end{itemize}

The combination of distributed control, heterogeneous autonomous systems, external interactions, and emergent behavior makes the smart warehouse a representative scenario for several contemporary challenges at the BPM-IoT intersection. In such dynamic settings, a complete and consistent view of the emergent process state and its context is neither available nor reconstructible. Therefore, interactions occurring under partial observability must be explicitly considered when modeling, operationalizing, and managing emergent IoT-enhanced processes at runtime. In the following, we elaborate on a subset of challenges, focusing on these modeling, operationalization, and runtime management aspects of emergence in IoT-enhanced business processes.

\section{Sub-Challenges and Problem Decomposition} \label{sec:subchallenges}

We divide the problem into three sub-challenges, SC1--SC3, aligned with the BPM lifecycle phases of process modeling, operationalization, and execution, since these phases are critical for managing uncertainty and emergent behavior \emph{at runtime}.

\subsection{SC1 -- Representing Uncertainty and Dynamics in IoT-Enhanced Processes}

In the \emph{modeling} phase of the BPM lifecycle, the emergent behavior of IoT-enhanced processes poses central challenges that are not fully addressed by current modeling approaches. We discuss this aspect by examining specific threats to effective process modeling induced by emergent behavior.

\begin{itemize}
    \item \emph{Uncertain event data}: Historical process data may be affected by uncertainties caused by several factors, both inherent and caused by emergence~\cite{pegoraro2019mining}. Sensor readings may be noisy, delayed, missing, duplicated, incorrectly attributed to cases, or ambiguous with respect to the activity they represent~\cite{franceschetti2023characterisation}. In this context, emergent behavior challenges the capabilities of established data-recording practices. As a result, process models need to represent not only events, but also confidence, provenance, ambiguity, and alternative interpretations; this is particularly crucial when analysts adopt data-driven modeling principles, such as process discovery.
    \item \emph{Unknown unknowns}: IoT-enhanced processes are shaped by ``unknown unknowns'', the parts of a process that are unobserved (and, at times, unobservable). Missing sensors, unobserved human workarounds, environmental conditions outside the measurement scope, and interactions between devices may influence the process without leaving a clean trace. This means that process intelligence cannot rely only on what is recorded, but must also reason about blind spots, incompleteness, and the limits of observability.
    \item \emph{Lack of semantics}: Devices, platforms, organizations, and physical contexts may each produce data according to their own local meanings, identifiers, time scales, and assumptions. When these environments interact, the same signal may have different implications, while similar events may not be semantically equivalent. IoT-enhanced business processes, therefore, need richer semantic mediation between local observations and global process understanding~\cite{bertrand2026nice,bertrand2025object}.
    \item \emph{Emergent drift:} IoT-enhanced processes may exhibit the gradual or sudden divergence between the intended process and the behavior produced by autonomous agents, automated devices, optimization routines, and human adaptation. Agent automation may improve local efficiency while unintentionally reshaping the overall process, creating new dependencies, shortcuts, risks, or coordination patterns.
\end{itemize}

\subsection{SC2 -- Operationalizing IoT-Enhanced Business Processes}

IoT-enhanced business processes operate in environments characterized by continuous data streams, heterogeneous actors, real-time interactions, and inherent unpredictability~\cite{BPIOTslr,valderas2022modelling}. Under these conditions, processes can no longer be treated as static and predefined control-flow structures. Instead, their operationalization is challenged by the need to account for unexpected runtime events, changing contextual information, and continuous interactions among process engines, IoT devices, software services, and external systems~\cite{elyasaf2025toward,BPIOTslr}.
We discuss the operationalization of IoT-enhanced business processes, focusing on the main concerns that arise when BPM systems are brought into direct interaction with IoT environments to make these processes executable.

\begin{itemize}
    \item \textit{Dynamic, event-driven, and context-aware execution}: Traditional process execution assumes that process models can be enacted according to predefined control-flow structures. In IoT-enhanced settings, however, execution is affected by runtime events, such as device unavailability, sensor observations, and physical-world interactions, which may trigger relevant context changes~\cite{schonig2018integrated}. 
    IoT-enhanced business processes may be driven by continuous streams of events produced by sensors, devices, platforms, and external services, while execution decisions may depend on environmental conditions, device states, user behavior, location, timing, or other situational factors~\cite{valderas2022modelling,poss2025location,vom2016role}. The challenge is that such events and contextual information may be frequent, heterogeneous, uncertain, delayed, missing, conflicting, distributed across different systems, or only available during execution. Process engines, therefore, have to cope with executions that may deviate from predefined assumptions and require runtime adjustment, while preserving process consistency and understandability.
    \item \textit{Interaction and architectural integration}: IoT-enhanced business processes often involve multiple IoT devices, actuators, platforms, services, and organizational systems~\cite{schonig2020iot}. A key challenge concerns how process engines interact with these heterogeneous components during execution. Such interaction may be \textit{passive}, when the process observes data streams and reacts to events produced by sensors or external systems; \textit{active}, when the process invokes devices, actuators, or IoT services during execution; or \textit{hybrid}, when both forms are combined within the same process. These interaction modes expose the limitations of isolated, monolithic, and ad hoc BPM solutions and architectures, raising challenges related to platform heterogeneity, communication protocols, reliability, latency, and coordination between the digital and physical worlds.
    This also requires managing dependencies among components that may evolve and fail independently, or interact in unforeseen ways during execution.
    \item \textit{Modeling, execution alignment, and reuse}: BPMN provides a rich and widely adopted notation for representing different types of process models, but its large set of notation elements and extensions also increases modeling complexity and requires clear usage practices~\cite{CompagnucciCFR24}. In IoT-enhanced settings, a key challenge is aligning what is modeled at design time with what can be executed at runtime, especially when processes depend on dynamic events, external devices, contextual information, and platform-specific capabilities~\cite{janiesch2020internet}. This challenge is intensified by the proliferation of solutions that are tied to specific domains or platforms, making it difficult to compare approaches, define reusable components (e.g., notation extensions), and build shared knowledge across systems. Open-source BPM systems are particularly relevant in this respect, but they also pose the challenge of defining standard, reusable extensions and shared practices to support IoT-driven process scenarios without becoming tied to specific use cases or platforms.
    \item \textit{Dynamic and mobile execution systems along the Cloud-Edge continuum}: In traditional process execution settings, the BPM systems are assumed to be centralized and statically placed on local or Cloud-based servers to facilitate client-server interactions and process executions involving multiple participants. For IoT-enhanced business processes, we see the need for more pervasive and mobile, personalized--\emph{ubiquitous}--BPM systems that are able to move along with the users--seamlessly transitioning between Cloud servers for resource-intensive and shared process executions and computations, to deployments on personal devices (e.g., smart phones) and closer to the edge for local sensor processing and actuations as part of the process executions~\cite{chang2016mobile}. Thereby, edge computing promises better support for real-time, safety, and privacy constraints through decentralized, local data processing and autonomous process execution, thereby reducing dependencies on constant, high-bandwidth Cloud connectivity~\cite{seiger2014distributed}. 
\end{itemize}

\subsection{SC3 -- Managing Emergent Behavior of Dynamic IoT-Enhanced Business Processes at Runtime}

We organize discussions on managing the emergent behavior of IoT-enhanced business processes during execution based on the \emph{MAPE-K} feedback loop proposed for autonomous computing~\cite{computing2006architectural}. This architecture has proven applicable for implementing autonomous behavior needed to deal with unanticipated situations, uncertainties, and emergence through self-adaptation across different system contexts~\cite{muccini2016self}.

\begin{itemize}
    \item \textit{Monitor}: Process and system monitoring has to be implemented in a traditional way within individual BPMS and IoT systems. Thereby, sensors can provide rich context information to contextualize process executions~\cite{bertrand2026nice}. However, when multiple BPM and IoT systems interact in unforeseen ways, monitoring needs to dynamically extend and adapt to multiple systems at runtime. This adaptation of monitoring is challenged by 1)~limited a priori knowledge regarding what specifically to monitor, 2)~changes in what to monitor at runtime, and 3)~constrained data collection capabilities (e.g., for privacy and security reasons). 
    \item \textit{Analyze}: During analysis, it is determined if system and process operations are within expected, normal parameters. This analysis usually assumes predetermined rules that define \emph{normal/desired} conditions and distinguish them from \emph{critical} conditions that require intervention. When dealing with emergence during process execution, we face uncertainty about how these analyses should be performed, since situations only emerge at runtime, when interventions might be needed (e.g., to avoid \emph{detrimental} emergent behavior~\cite{kopetz2016emergence}). Data analysis at runtime might require the dynamic synthesis and adaptation of analysis rules, as well as accepting incomplete knowledge as given to cope with changing situations and uncertainties.
    \item \textit{Plan}: Planning is responsible for deriving an adaptation plan to change the current systems and process executions based on change requests from the analysis, bringing the system (of systems) operations back into a desired state. Thereby, planning is challenged by uncertainties about what constitutes a desired state — knowledge that may be only partially available at design time. At runtime, novel, unanticipated situations might emerge under an open-world assumption, which can push state-of-the-art planning approaches to their limits.   
    \item \textit{Execute}: Execution carries out the action plan derived during planning to move operations into a desired state again. In highly complex SoS scenarios where constituent systems influence each other in unanticipated ways, the execution of each corrective action might lead to unforeseen effects on other systems and their environment. The execution is challenged by continuously changing local and global system states and contexts, as well as newly emerging ones that require constant, high-frequency situation analysis, evaluation, and re-planning. Still, execution has to maintain real-time and safety guarantees in these highly dynamic settings as we assume that the execution of a corrective action might also involve manipulating the physical world via actuators.  
    \item \textit{Knowledge}: The knowledge base contains all relevant information and data to enable the autonomous behavior as each of the four phases constantly consults and updates this knowledge during operations. Creating and maintaining this knowledge base is challenging, as we constantly face uncertainties and incomplete knowledge when novel situations arise. During process execution, knowledge might be available only locally to the responsible BPM system and the involved IoT systems, and be constrained to their execution contexts. Other systems might operate in parallel and in their vicinity, but knowledge is incomplete and not shared for different reasons (e.g., privacy or unanticipated situations). Larger, new execution systems and contexts might emerge, but it is unclear how the knowledge should be managed, shared, and distributed.
\end{itemize}

\section{State of the Art} \label{sec:sota}

\paragraph{Representing Uncertainty and Dynamics in IoT-Enhanced Processes.}
Several approaches incorporate contextual information into IoT-enhanced execution by using ontologies, high-level events, or context models derived from IoT data to influence process behavior at runtime~\cite{bertrand2026nice,poss2025location,valderas2022modelling,ochoa2024dynamic}. 
Nevertheless, context is often treated as external information consumed by the process rather than as a first-class construct explicitly represented and managed within the execution model. 

In some lines of work, context carries meta-information about data uncertainty itself. When available, information related to uncertainty can be leveraged to obtain more trustworthy process analysis~\cite{pegoraro2022probabilistic}. Various post-mortem analysis methods for metadata-augmented event logs are available, including conformance checking and process discovery~\cite{pegoraro2021proved}. While this paradigm can address data uncertainty, the available techniques are foundational, and their application to IoT is in its infancy. Early work includes process querying in settings with anomalous timestamps~\cite{busany2020interval}. Other approaches focus on \emph{repairing} the data, rather than analyzing it natively: several error-correction techniques are available, generally model- or rule-based, for both data correction and imputation~\cite{shirali2024interactive,seeliger2024inferring}.

\emph{Ambiguity} stemming from IoT data processing, a specific source of uncertainty in BPM, is discussed by the authors in~\cite{franceschetti2025proambition}. The authors argue that there is often a gap between the low-level data streams emitted by IoT devices and the associated business process data. Performing event abstraction of IoT data to close this gap might lead to ambiguities for several reasons~\cite{franceschetti2023characterisation} and requires an explicit representation of identified uncertainties for subsequent process analytics~\cite{franceschetti2025toward}.

In summary, recent survey work identifies preprocessing, activity recognition/discovery, event abstraction, case identification, event-activity correlation, and event ordering as recurring steps in analysis~\cite{brzychczy2025process,oukharijane2018survey}, which help address anomalous IoT data recordings. However, there is still limited work on a unified framework that represents IoT data uncertainty explicitly and carries it through repair, event-log construction, and downstream process-mining analysis.

\paragraph{Operationalizing IoT-Enhanced Business Processes.} 
Research on the operationalization of IoT-enhanced business processes has initially addressed the problem from a modeling perspective, focusing on how IoT entities, data, events, and physical-world interactions can be represented within process models. 
Early contributions mainly focused on extending process modeling languages to represent IoT-related elements and physical-world interactions. For instance,~\cite{Meyer2013IoTAwareProcessModeling} extends BPMN with sensing and actuation tasks, physical entities, and real-world data objects, while~\cite{uBPMN} introduces dedicated tasks for ubiquitous process interactions, including sensors, actuators, readers, and collectors. Other approaches focus on more specific IoT execution settings:~\cite{sungur2013bpmn4wsn} focuses on operations executed by wireless sensor network devices, whereas~\cite{appel2014event} introduces constructs for specifying and processing event streams produced by smart sensors. These works make IoT interactions explicit at design time and help reduce the gap between process models and their runtime interpretation.

However, modeling IoT elements is only one part of operationalization. IoT-enhanced processes must also be enacted in environments characterized by continuous data streams, heterogeneous devices, and changing contextual conditions. This shift toward execution is reflected in recent evidence: among 84 analyzed approaches, 43 focus only on design, whereas 41 also address enactment or execution~\cite{BPIOTslr}. This indicates that operationalization is now recognized as a central concern, although the relation between design-time process models and executable IoT-aware systems remains only partially consolidated. 

Given the event-driven nature of IoT-enhanced execution, several approaches have been proposed to support runtime event processing and event integration~\cite{kirikkayis2023integrating}. In these approaches, BPM engines are integrated with Complex Event Processing (CEP), stream processing, or event processing components to enable process instances to respond to real-time IoT data streams~\cite{seiger2022integrating,valderas2022modelling}. Other contributions consider how contextual information, such as location, device states, or environmental data, can influence process execution and monitoring~\cite{poss2025location}. While these works support more responsive and data-driven execution, event and context handling is typically delegated to external components, highlighting the limited native support for IoT-aware execution in BPM engines.

Further work focuses on the architectural integration between BPM systems and IoT environments~\cite{chang2016mobile}. Existing solutions include service- and microservice-based approaches that abstract IoT devices as services~\cite{valderas2022modelling}, end-to-end frameworks that extend BPMN and integrate IoT-aware execution components~\cite{kirikkayis2023bpmne4iot,poss2025location}, and BPMN-driven frameworks for cyber-physical or robotic systems~\cite{corradini2023bpmn}. While these works all assume centralized BPM systems, the authors in~\cite{seiger2017self} also discuss distributed process executions in IoT scenarios (e.g., involving robots) that require additional self-adaptation capabilities. These approaches demonstrate the feasibility of operationalizing IoT-enhanced processes, but they usually rely on framework-specific assumptions, execution architectures, and integration mechanisms.

As a consequence, the state of the art still lacks shared practices and reusable extension mechanisms to support the systematic operationalization of IoT-enhanced business processes across heterogeneous IoT settings. In particular, existing approaches offer limited support for migrating from IoT-aware process models to reusable, portable, and runtime-aware execution infrastructures capable of handling heterogeneous devices, dynamic events, contextual changes, and execution-time uncertainty.

\paragraph{Managing emergent behavior of dynamic IoT-enhanced business processes at runtime.}
Several existing contributions focus on adaptive business processes and BPM systems in traditional (non-IoT related) settings, e.g., with a focus on context adaptation~\cite{hermosillo2010creating} and self-adaptation mechanisms~\cite{oukharijane2018survey}. IoT-enhanced business processes do not play a significant role in these works.
More IoT/CPS-related approaches investigate how running process instances can react to contextual changes, exceptional situations, or sensor-detected events, for instance, through feedback-control mechanisms such as MAPE-K loops~\cite{seiger2019toward,malburg2023applying} or through broader BPM-IoT adaptation perspectives in centralized process executions~\cite{domingos2010ad,leotta2019iot,marrella2016intelligent}, but also in decentralized, pervasive execution scenarios~\cite{seiger2017self,li2026cargo}. 
However, adaptation is often addressed at the architectural level or through external mechanisms, rather than being explicitly governed within the BPM execution layer. Furthermore, these approaches focus on individual, self-contained IoT system contexts and local feedback loops. More complex SoS configurations and emergent behavior that require decentralized, dynamic--and potentially adaptive--self-adaptation mechanisms are not discussed.

Emergent behavior in connection with business processes, e.g., using process mining for emergent behavior analysis, is discussed in~\cite{bemthuis2019agent}. In a follow-up work, the authors propose an agent-based simulator for the detection of emergent behavior in CPS~\cite{bemthuis2020using}. While this research is highly relevant for observing and detecting emergent behavior in IoT-enhanced business processes within the monitor and analysis phases of the feedback loops, they do not extend to the reactive part, and they do not discuss runtime capabilities. On the other hand, approaches to address and enact emergent behavior in IoT systems (e.g., via \emph{emergent configurations}~\cite{alkhabbas2017architecting}, domain objects~\cite{alkhabbas2018enacting} or \emph{Fog Computing}~\cite{roca2017tackling}) exist, but they do not link to (business) processes. A comprehensive investigation of emergent behavior in cyber-physical systems of systems can be found in~\cite{kopetz2016emergence}. This work constitutes a fundamental pillar for research on emergence in SoS and IoT. However, the link to IoT-enhanced \emph{business processes} still needs to be created.

\section{Gaps and Opportunities} \label{sec:gapsopportunities}

Building on insights from state-of-the-art investigations, we identify current research gaps and propose novel research directions.

\paragraph{From predefined process models to runtime emergence:} 
Traditional BPM approaches rely on predefined process models, assuming that relevant execution paths can be anticipated at design time.
In IoT-enhanced environments, this assumption no longer holds. In such contexts, multiple sources of uncertainty arise: devices may fail, human workers may deviate from expected behavior, environmental conditions may change, and unforeseen events may occur. These \textit{unknown unknowns} generate emergent behaviors that can not be fully anticipated during process modeling~\cite{perez2014uncertainties,filippone2024handling}.
Despite prior work on adaptive and flexible processes, existing BPM engines still typically require fully specified models and offer limited support for partial process definitions, dynamic composition, or runtime emergence. Accordingly, a key opportunity is to define modeling approaches that make uncertainty explicit, together with execution mechanisms that can support emergent behavior while preserving process consistency and understandability.

\paragraph{Uncertainty as a first-class citizen in self-adaptive systems:}
Uncertainties are ubiquitous in SoS contexts and we assume that it is rarely possible to resolve all uncertainties. Thus, uncertainty needs to be made explicit throughout all considered phases of the BPM lifecycle and, specifically, in the MAPE-K architecture.
Accordingly, a key opportunity is to make uncertainty explicit not only within the BPM lifecycle but also across the MAPE-K phases, so that monitoring, analysis, planning, execution, and knowledge management can reason under incomplete or missing information.

\paragraph{Context as a first-class citizen:} 
In IoT-enhanced processes, context not only describes the execution environment. It can also trigger, constrain, and redirect process behavior. Location, device state, environmental conditions, resource availability, and human-related factors may determine which activities are enabled, how decisions are made, and when adaptations are required.
Yet, context is often treated as external data consumed by the process, rather than as a first-class element that shapes execution semantics and runtime behavior. Accordingly, a key opportunity is to define context models that explicitly capture the relation between contextual conditions and process behavior, including context validity, uncertainty, and impact on adaptation.

\paragraph{Adoption of unified ontologies and taxonomies:} Shared ontologies and taxonomies provide a common vocabulary for linking low-level sensor data, contextual information, process concepts, and behavioral models, and thus are essential for analyzing IoT processes affected by uncertainty and emergent behavior~\cite{bertrand2025object}. Data-oriented taxonomies help characterize quality issues such as missing, ambiguous, noisy, or imprecise events, making assumptions explicit and comparable across studies. Model-oriented ontologies support the representation of activities, resources, objects, constraints, and expected interactions, enabling uncertain observations to be interpreted using domain knowledge rather than discarded. Together, they improve interoperability, support reproducible event-log construction, guide repair and abstraction decisions, and enable distinguishing true process deviations from artifacts of sensing, correlation, or modeling limitations.

\paragraph{From discrete events to continuous event streams:}
BPM engines traditionally react to discrete events such as messages, timers, or service responses. In IoT-enhanced environments, execution may depend on continuous, high-frequency, heterogeneous, and uncertain event streams produced by sensors, devices, platforms, and external actors.
Although CEP and stream-processing technologies are available, their integration with BPM engines is often externalized and architecture-specific~\cite{schonig2020iot}. A key opportunity is to clarify how event processing and process execution should interact, including event-process correlation, stream-aware execution, and scalability under continuous IoT data.

\paragraph{Distributed BPM systems for ubiquitous process executions:} Moving away from centralized client-server architectures, BPM systems have to become more modular, distributed, and mobile to support decentralized and personalized executions of IoT-enhanced business processes. These process executions should facilitate migration between low-footprint, pervasive, and portable process engines on embedded edge devices (e.g., in containerized environments or WebAssembly modules), personal devices (e.g., smartphones and laptop computers), and highly distributed and scalable Cloud resources. This requires the development of 1)~workflow engines that are able to operate in resource-constrained environments (e.g,~on robots~\cite{seiger2026autonomous}) and 2)~interoperable solutions that allow moving process instances between different types of BPMS provided and operated by and on different entities.

\paragraph{Continuous, emerging, high-frequency adaptation loops with safety constraints:}
In IoT settings and SoS, operations and context are constantly changing, creating unanticipated situations that need to be reacted and adapted to via feedback loops (e.g., implementing MAPE-K) at runtime. These feedback loops have to be used to manage all involved components and processes and they might have to be dynamically synthesized and adapted at runtime. Despite the flexibility and potential overhead introduced by these additional control loops, safety and real-time constraints imposed by IoT systems must be met. The implementation of AI-based agents appears to be a promising direction to realize the self-adaptation mechanisms, as they are capable of advanced reasoning for runtime decision making using large knowledge bases and world models. However, they cannot meet the current safety requirements of CPS/IoT systems, which require real-time decision-making and action. 

\paragraph{From local to ad-hoc, global context models and knowledge:}
The knowledge base is an important pillar for implementing feedback loops to manage the emergent behavior of IoT-enhanced business processes at runtime. So far, knowledge is available only locally within the individual constituent systems. In emerging interactions in SoS scenarios, the knowledge bases have to be created and extended at runtime to allow for knowledge sharing among systems in order to agree upon a common \emph{world model}, which could, for example, be based on a multi-faceted \emph{digital twin} of the relevant systems and their business processes~\cite{fornari2025digital,DPT}. However, privacy and security constraints of data imposed upon by individual systems have to be respected.

\paragraph{From ad-hoc adaptation to governed runtime adaptation:} 
IoT-enhanced business processes must adapt to continuous context changes and event streams. This can be framed as a form of \textit{self-modification}, where the presented gap concerns runtime \textit{adaptation} rather than longer-term process \textit{evolution}~\cite{elyasaf2025toward}.
Runtime adaptation introduces risks related to process consistency, constraint violations, and unintended behavior. 
Although some engines and solutions support adaptability, adaptation is often ad hoc and context-dependent.
Accordingly, a key opportunity is to define governed adaptation mechanisms that support controlled runtime changes, including adaptation policies, verification, rollback, recovery, and principles for deciding when human oversight is needed.

\paragraph{Meta-adaptations in highly dynamic System-of-Systems contexts:}
Currently, MAPE-K-based mechanisms for self-adaptation rely on static approaches to monitoring, analysis, planning, and execution, which work well in closed systems. When aiming to manage emergent behavior in SoS, these approaches may also have to become more decentralized, dynamic, and self-adaptive~\cite{de2013software}, i.e., each of the MAPE phases managing a system or a process might itself be managed by a feedback loop, leading to \emph{meta-adaptations}. By treating the MAPE-based feedback loops as partial business processes themselves, these meta-adaptations can be realized in a unified, process-driven way, using the same process syntax and execution systems as the business processes or fragments managed by these control loops.

\paragraph{Shared architectures and practices:}
Existing BPM-IoT solutions often rely on modeling- and engine-specific extensions, platform-specific integrations, or domain-specific architectures. While these solutions demonstrate feasibility, they remain difficult to compare, reuse, and transfer across settings.
Accordingly, a key opportunity is to define shared extension mechanisms, reusable integration patterns, and open reference architectures that clarify how BPM engines, IoT middleware, event processing, context management, and adaptation logic should interact.

\section{Recommendations} \label{sec:recommendations}

Based on the identified gaps and opportunities, we recommend the following action items for future, short-term, medium-term, and long-term research and development activities. With every recommendation, we associate the specific sub-challenges (SC) introduced in Section~\ref{sec:subchallenges} that it addresses.

\subsection{Short-Term Recommendations}

In the short term, research should focus on clarifying the assumptions and technical foundations needed to study emergent IoT-enhanced business processes. First, existing techniques for uncertainty-aware event data recording and publishing should be implemented and assessed, while uncertainty should also be investigated across the phases of MAPE-K feedback loops. This supports the representation and propagation of uncertainty across modeling and runtime management \textbf{[SC1, SC3]}.

Second, existing BPM engines should be analyzed and benchmarked for IoT-driven and event-based execution. This includes clarifying how BPM engines can interact natively with event-processing technologies, such as CEP and stream-processing platforms, and assessing the suitability and resource impact of current BPM systems for edge deployments \textbf{[SC2]}.

Third, future work should identify key sources of unpredictability and define initial first-class context models. At the same time, the implications of incomplete, local, and evolving knowledge for runtime management should be analyzed, including the role of digital twins in maintaining up-to-date knowledge bases that can be integrated while preserving privacy \textbf{[SC1, SC2, SC3]}.

Finally, requirements for runtime monitoring and adaptation to emergent behavior in IoT-enhanced business processes operating in continuously evolving execution contexts should be identified~\textbf{[SC3]}.

\subsection{Medium-Term Recommendations}

In the medium term, research should move from assessment toward the development of concrete mechanisms. For uncertainty-aware modeling and analysis, process analysis techniques should be adopted that account for non-deterministic and probabilistic information, while also relaxing assumptions such as case-based and total-order event structures \textbf{[SC1]}.

At the execution level, new mechanisms are needed to handle uncertainty and emergent behavior at runtime. This includes defining explicit runtime adaptation policies, constraint-preservation mechanisms, and rollback strategies, as well as establishing structured approaches for integrating context models into execution logic \textbf{[SC1, SC2, SC3]}.

To improve operationalization, standardized extension points should be introduced at both modeling and execution levels. In parallel, low-footprint and standard-compatible BPM systems should be developed to support process execution on resource-constrained devices \textbf{[SC2]}.

For runtime management, MAPE-K feedback loops should be integrated as process-driven (\emph{process-native}) concepts into IoT-enhanced business processes. This requires decentralized runtime monitoring mechanisms and privacy- and goal-oriented approaches for knowledge disclosure and distributed world-model construction across interacting systems \textbf{[SC3]}.

\subsection{Long-Term Recommendations}

In the long term, the community should work toward shared practices for emergent IoT-enhanced BPM. This includes rigorously defined semantics, taxonomies, and ontologies for process components, actors, and variables, as well as techniques for estimating emergent drift caused by automation, limited environmental control, and external agents \textbf{[SC1]}.

At the architectural level, scalable BPM architectures for IoT environments should be designed and validated. These architectures should integrate BPM engines, IoT middleware, event processing, adaptation mechanisms, and reusable integration patterns. They should also support the seamless migration of BPM systems and process executions along the Cloud-Edge continuum \textbf{[SC2]}.

The community should also promote reusable components and shared practices across BPM and IoT ecosystems, enabling adaptive, context-aware processes that can evolve continuously \textbf{[SC2, SC3]}.

Finally, runtime management should evolve toward distributed and meta-adaptive infrastructures. This includes distributed knowledge for large-scale emergent IoT-enhanced SoS, (AI) agent-based feedback-loop mechanisms with real-time decision-making and actuation guarantees, and meta-adaptive MAPE-K architectures capable of modifying their own feedback-loop mechanisms at runtime \textbf{[SC3]}.

\section{Conclusions} \label{sec:conclusions} 
IoT-enhanced business processes increasingly operate in dynamic environments characterized by heterogeneous actors, autonomous systems, and evolving execution contexts. In such settings, process behavior cannot be fully anticipated as it emerges through complex interactions among humans, IoT devices, and (possibly external) software systems. Emergent behavior in such system-of-systems contexts is characterized by partial observability and uncertainty, which challenge traditional BPM assumptions. In this paper, we discussed these challenges from three complementary perspectives grounded in the BPM lifecycle, addressing uncertainty representation, the operationalization of IoT-enhanced process execution, and the runtime management of emergent behavior during execution. Our analysis of the state of the art identified several research gaps. Based on these gaps, we proposed short-, medium-, and long-term recommendations to stimulate further research and development toward more adaptive, context-aware, interoperable approaches to IoT-enhanced business processes. Overall, we argue that effective management of emergent behavior requires shifting from static, centrally controlled processes toward self-adaptive, decentralized, and uncertainty-aware approaches that can cope with continuously evolving execution contexts, supported by shared practices. Future research should, therefore, focus on uncertainty-aware approaches to modeling, execution, and runtime management; distributed knowledge and world-model construction; decentralized, self-adaptive mechanisms and executions; autonomous and interoperable architectures and shared practices. The presented challenges, gaps, and recommendations contribute toward establishing a future research agenda at the intersection of BPM, IoT, and emergent SoS.

\begin{acknowledgments}
  This work has been partially funded by the MUR (Italy) Department of Excellence 2023 - 2027 for GSSI. This work has received funding from the Swiss National Science Foundation under Grant No. 10002384 (\emph{TaSSAreCt} project). This work has been supported by the Internet of Processes and Things (IoPT) community through discussions, datasets, and feedback.
\end{acknowledgments}

\section*{Declaration on Generative AI}
During the preparation of this work, the authors used GPT-5.5 and Grammarly for grammar and spell checks. Figure 1 was also AI-generated, as stated in its caption.
After using these tools/services, the authors reviewed and edited the content as needed and took full responsibility for it.

\bibliography{bib}

\end{document}